\documentclass{aa}  

\usepackage{graphicx}
\usepackage{txfonts}
\usepackage{xspace}
\usepackage{color}
\usepackage{url}
\usepackage{hyperref}
\usepackage{siunitx}
\newcommand{\Nu}{{\it NuSTAR\xspace}}
\newcommand{\ch}{{\it Chandra\xspace}}
\newcommand{\xm}{{\it XMM-Newton\xspace}}

\newcommand{\sw}{{\it Swift\xspace}}
\newcommand{\source}{2S 1845$-$024\xspace}

\usepackage[flushleft]{threeparttable}
\usepackage{graphicx,epstopdf}
\DeclareGraphicsExtensions{.ps}
\def\ergscm{erg~s$^{-1}$~cm$^{-2}$}

\def\flux{erg s$^{-1}$ cm$^{-2}$}

\begin{document}

\title{Broad-band analysis of X-ray pulsar 2S 1845$-$024}

\author{Armin Nabizadeh \inst{1} 
          \and Sergey S. Tsygankov \inst{1,2}
   		  \and Sergey V. Molkov \inst{2}
   		  \and Dmitri I. Karasev \inst{2}
   		  \and Long Ji \inst{3}
          \and Alexander~A.~Lutovinov\inst{2}
   		  \and Juri Poutanen \inst{1,2,4}
          }
          
   \institute{Department of Physics and Astronomy, FI-20014 University of Turku,  Finland \\ \email{armin.nabizadeh@utu.fi}
       \and
       Space Research Institute of the Russian Academy of Sciences, Profsoyuznaya Str. 84/32, Moscow 117997, Russia    
       \and
       School of physics and astronomy, Sun Yat-Sen University, Zhuhai, Guangdong 519082, China
       \and
       Nordita, KTH Royal Institute of Technology and Stockholm University, SE-10691 Stockholm, Sweden
          }
          
\titlerunning{Properties of XRP 2S 1845$-$024}
\authorrunning{Nabizadeh et al.}

   \date{2021}

\abstract{We present results of detailed investigation of the poorly studied X-ray pulsar \source\ based on the data obtained with \Nu\ observatory during the type I outburst in 2017. Neither pulse phase-averaged, nor phase-resolved spectra of the source show evidence for a cyclotron absorption feature. We also used the data obtained from other X-ray observatories (\sw, \xm\ and \ch) to study the spectral properties as a function of orbital phase. The analysis revealed a high hydrogen column density for the source reaching $\sim$10$^{24}$~cm$^{-2}$ around the periastron. Using high-quality \ch\ data we were able to obtain an accurate localization of \source\ at R.A. = 18$^{\rm h}48^{\rm m}16\fs8$ and Dec. = $-2\degr25\arcmin25\farcs1$ (J2000) that allowed us to use infrared (IR) data to  roughly classify the optical counterpart of the source as an OB supergiant at the distance of $\gtrsim$15 kpc.
}


\maketitle
%
\section{Introduction}

\source (also known as GS 1843$-$024) is a transient X-ray source discovered with the {\it Ginga} observatory \citep{1988IAUC.4661....2M,Koyama1990}. It belongs to the class of high mass X-ray binaries (HXMB). Many of the physical properties of the system and the neutron star (NS) are still unknown. The system contains an X-ray pulsar (XRP) with a spin period $P_{\rm spin}$ = 94.8 s \citep{Makino1988,Zhang1996}. A series of the Burst and Transient Source Experiment (BATSE) observations performed in 1991--1997 detected 10 type I outbursts revealing an orbital period $P_{\rm orbital}$ =  242.18 $\pm$ 0.01~d for the system \citep{Finger1999}. More outbursts around periastron passage (orbital phase zero) were detected later by different observatories \citep[e.g.,][]{2008ARep...52..138D}. No type II outbursts have yet been detected from the source. The timing analysis allowes to determine the orbital parameters of the system: the high eccentricity of $e=0.879\pm0.005$ and the projected semi-major axis $a_{\rm x}\sin i = 689 \pm 38$ lt-s, suggesting a high-mass companion (M > 7$M_{\rm \odot}$) for \source\  \citep{Finger1999,Koyama1990}.

The companion star in this system has not yet been directly identified. However, the source is classified as a transient Be/XRP based on the outburst pattern and the highly eccentric orbit \citep{Koyama1990,Zhang1996,Finger1999}. In addition, the location of the source in the \citet{Corbet1986} diagram is consistent with a Be/NS binary. The 2--38 keV X-ray spectrum of \source\, obtained by the {\it Ginga} Large Area Counter (LAC), fitted by a power-law with a high-energy cutoff model, revealed a large hydrogen column density $N_{\rm H}$ $\simeq$ $(1.5-3.0) \times 10^{23}$ cm$^{-2}$ in the direction to the source \citep{Koyama1990}. Assuming that the lower limit on $N_{\rm H}$ is accounted for by the interstellar medium, \citet{Koyama1990} estimated the source distance to be about 10 kpc. We emphasize that there is no {\it Gaia} distance measurements available for this source.
 
The BATSE observations of \source\ also measured a secular long-term spin-up trend at a rate of $\dot{\nu} \sim 2.7 \times 10^{-13}$ Hz s$^{-1}$ during the 1991--1997 period of activity \citep{Finger1999}. Currently, however, the observations provided with the {\it Fermi} Gamma-ray Burst Monitor (GBM) Accreting Pulsars Program \citep[GAPP\footnote{\url{http://gammaray.nsstc.nasa.gov/gbm/science/pulsars/}};][]{Malacaria2020} show that the source has been in a spin-down phase during the last six years. It can be, therefore, inferred that the source had undergone a torque reversal before entering to the long-term spin-down trend with a rate $\dot{\nu} \sim -2.4 \times 10^{-13}$ Hz s$^{-1}$ \citep{Malacaria2020}. Because there is no data available for the source in the period between 51560 and 56154 MJD, \cite{Malacaria2020} estimated the torque reversal occurred on 53053 $\pm$ 250 MJD by extrapolating the spin-up and spin-down log-term trends in the gap between BATSE and GBM observations.

Although, there are several X-ray observations available for \source, the properties of the source in the soft and hard X-ray bands have not been fully investigated. Namely, some fundamental parameters as the NS magnetic field strength, the type of the companion star, and the distance to the system are not determined or still under debate. In the current work, we used a single \Nu\ observation, which was performed during a normal type I outburst on 2017 April 14 as well as several other archival observations obtained with different X-ray satellites, to perform a detailed temporal and spectral analysis of \source\ in a wide energy band in order to determine its properties.

\section{Observations and data reduction}

\begin{figure}
\begin{center} 
\includegraphics[width=0.9\columnwidth]{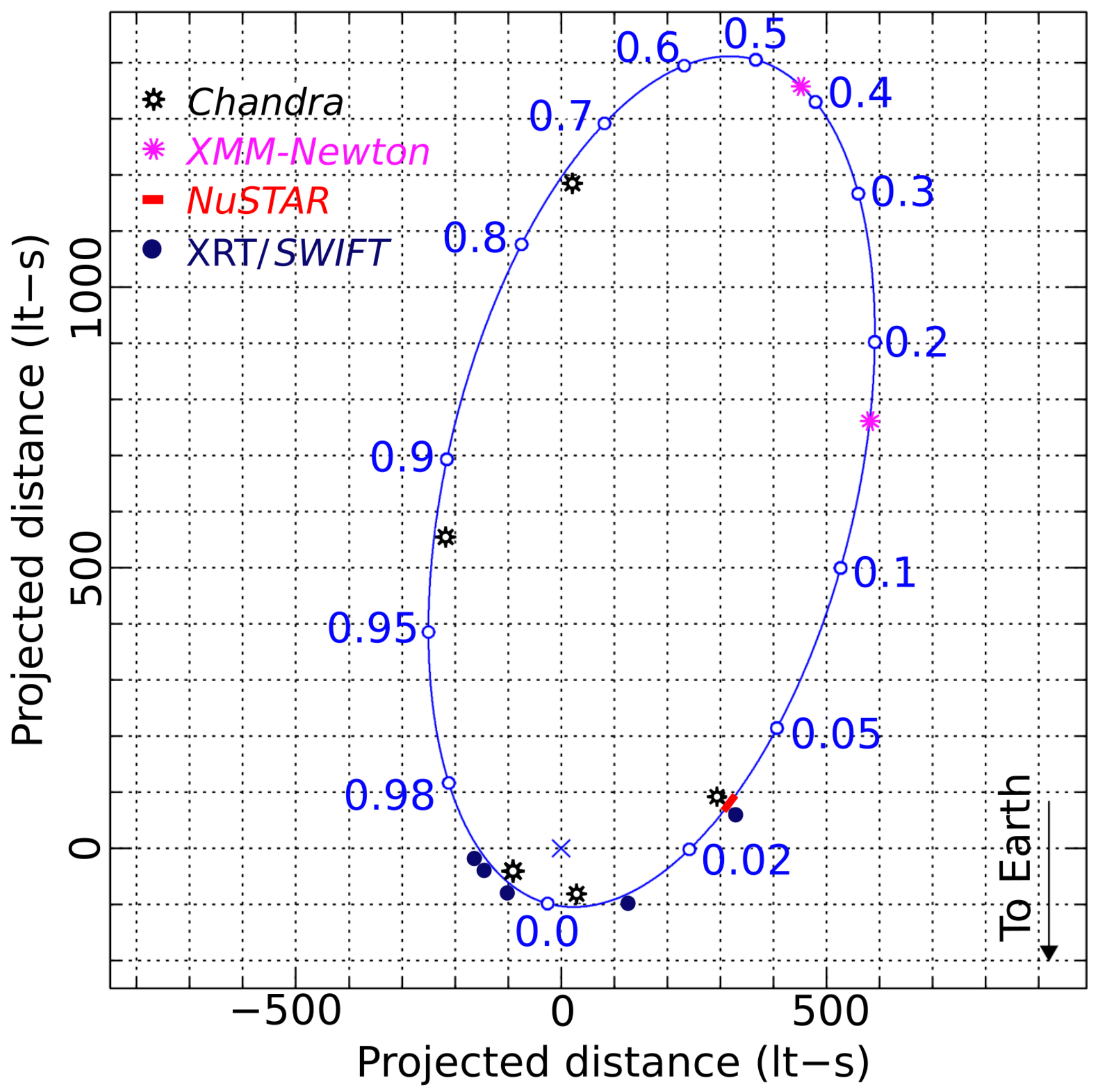}
\end{center}
\caption{Orbital phases corresponding to the date of each observation performed by \Nu, \xm, \ch\ and \sw/XRT.}
\label{fig:orbit} 
\end{figure}

Since the discovery, \source\, has been extensively observed by several instruments such as \Nu, \xm, \ch\ and \sw. The summary of the observations utilized in our work is given in Table~\ref{tab:observations}. Here we focus on the details of the observations obtained by the mentioned X-ray observatories which were performed at different orbital phases (see Fig.~\ref{fig:orbit}) calculated using ephemeris $T_{\rm Periastron}$ = 2449616.98$\pm$0.18 (JD) \citep{Finger1999}. The temporal and spectral analysis was done using {\sc heasoft} 6.28\footnote{\url{http://heasarc.nasa.gov/lheasoft/}} and {\sc xspec} 12.11.1b\footnote{\url{https://heasarc.gsfc.nasa.gov/xanadu/xspec/manual/XspecManual.html}}. For the spectral analysis the data were grouped to have at least 25 count per energy bin in order to use $\chi^2$ statistics unless otherwise stated in the text.

\subsection{\Nu\ observations}
\label{nu-observations}

\Nu\ X-ray observatory consists of two identical and independent co-aligned X-ray telescopes focusing the incident X-rays into two Focal Plane Modules A and B (FPMA and FPMB) \citep{Harrison2013}. The instruments contain four (2$\times$2) solid-state cadmium zinc telluride (CdZnTe) pixel detectors operating in a wide energy range of 3--79 keV. \Nu\ instruments provide an X-ray imaging resolution of 18$\arcsec$ full width at half maximum (FWHM) and a spectral energy resolution of 400 eV (FWHM) at 10 keV. \source\ was observed with \Nu\ on 2017 April 14 for a duration of $\sim$35 ks during the peak of the outburst. In order to reduce the raw data, we followed the standard procedure explained in \Nu\ official user guides\footnote{\url{https://nustar.ssdc.asi.it/news.php\#}} the \Nu\ Data Analysis Software {\sc nustardas} v2.0.0 with a {\sc caldb} version 20201130. The source and background photons were extracted from circular regions with radii 90$\arcsec$ and 150$\arcsec$, respectively, for both the modules.

\subsection{\sw\ observations}
\label{swiftdata}

\source\ was observed by the XRT telescope \citep{Burrows2005swift} onboard the \textit{Neil Gehrels Swift Observatory} \citep[\sw;][]{Gehrels2004swift} several times in the period of 2007--2019. In this study, we used five \sw/XRT observations, all obtained in the photon counting (PC) mode as listed in Table~\ref{tab:observations}. The corresponding spectra were extracted using the online tools\footnote{\url{https://www.swift.ac.uk/user_objects/}} \citep{Evans2009XRTonline} provided by the UK Swift Science Data Centre. Because the count rate in all \sw\ observations is below 0.3 count s$^{-1}$, the data were not affected by the pile-up effect.\footnote{\url{https://www.swift.ac.uk/analysis/xrt/pileup.php}}
One of the \sw/XRT observations (ObsID 00088089001) were performed simultaneously with the \Nu\ observation allowing us to get the spectral parameters in a wider energy band 0.3--79 keV. The source spectra as observed by \sw/XRT and \Nu/FPMA-B were then fitted simultaneously in the energy range 0.3--10 and 4--79 keV, respectively, accounting for difference in a normalization.

\begin{table}
    \centering
    \caption{Observation log of \source.}
    \begin{tabular}{cccc}
    \hline \hline
    ObsID & Start date & Start MJD & Exposure (ks) \\
    \hline
    \multicolumn{4}{c}{\Nu}  \\
    90201056002 & 2017-04-14 & 57857.59 & 34.71 \\
    \multicolumn{4}{c}{\xm}  \\
    0302970601 & 2006-04-11 & 53836.75 & 22.66 \\
    0302970801 & 2006-10-06 & 54014.39 & 15.91 \\
    \multicolumn{4}{c}{\ch}  \\
    2692 & 2002-08-18 & 52504.25 & 4.96  \\
    2689 & 2002-09-04 & 52521.42 & 14.80 \\
    2691 & 2002-09-06 & 52523.31 & 14.76 \\
    2690 & 2002-09-12 & 52529.78 & 15.09 \\
    10512 & 2009-02-21 & 54883.40 & 5.76 \\
    \multicolumn{4}{c}{\sw/XRT}  \\
    00609139000 & 2014-08-10 & 56879.59 & 0.80 \\
    00033739001 & 2015-04-14 & 57126.04 & 0.59 \\
    00707545000 & 2016-08-06 & 57606.47 & 1.53 \\
    00745966000 & 2017-04-06 & 57849.52 & 0.57 \\
    00088089001 & 2017-04-14 & 57857.82 & 1.98 \\
    \multicolumn{4}{c}{\it UKIDSS/UKIRT}  \\
     4543927 & 2006-06-12 & 53898.468 & 0.39 \\
     6610544 & 2006-06-12 & 53898.472 & 0.36\\
    \hline
    \end{tabular}
    \label{tab:observations}
\end{table}

\begin{figure}
\begin{center} 
\includegraphics[width=0.9\columnwidth]{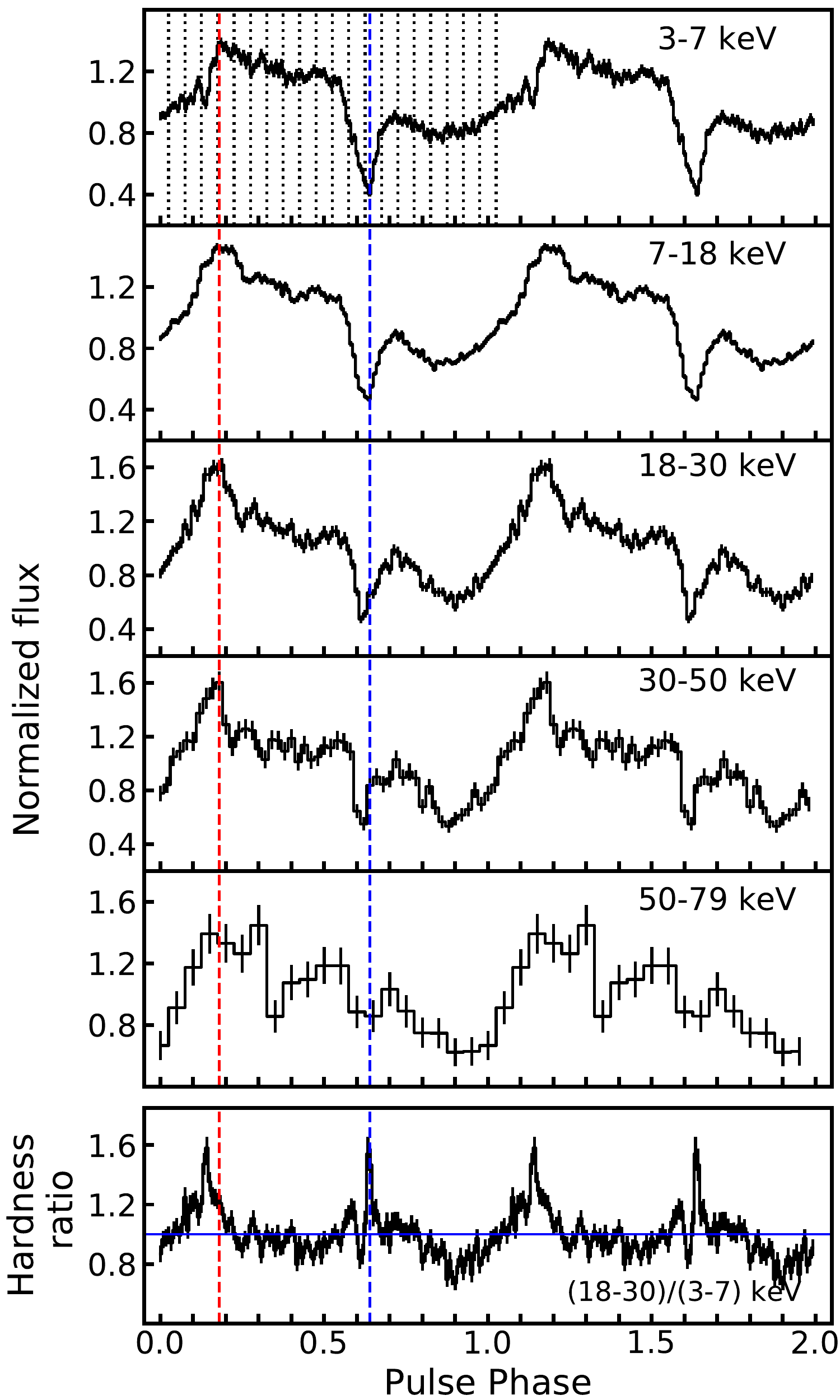}
\end{center}
\caption{\textit{Top panels:} Pulse profile of \source\ in different energy bands obtained from the \Nu\ observation. Fluxes are normalized by the mean flux in each energy range. The red and blue dashed lines show the main maximum and minimum in the 3--7 keV band, respectively. The black dotted lines in the most upper panel show the phase segments which were used to extract the phase-resolved spectra. \textit{Bottom panel:} Hardness ratio of the source over the pulse phase calculated as a ratio of normalized count rates in the pulse profiles in the energy bands 18--30 and 3--7 keV. 
The hardness ratio of the unity is indicated by the horizontal blue solid line.}
\label{fig:PP} 
\end{figure}

\subsection{\xm\ observations}
\label{xmmtData}

The X-ray Multi-Mirror Mission (\xm) \citep{Jansen2001} carries three X-ray telescopes each with a medium spectral resolution European Photon Imaging Camera at the focus operating in the range of 0.2--10 keV (EPIC-MOS1, -MOS2 and -pn). 
\source\ was observed by \xm\ two times in 2006 with the exposure times of $\sim$23 and $\sim$16 ks with all three EPIC X-ray instruments. We reduced and analyzed the data following the standard procedure explained in Science Analysis System (SAS) user guide\footnote{\url{https://xmm-tools.cosmos.esa.int/external/xmm_user_support/documentation/sas_usg/USG/}} using the software SAS version 17.0.0 and the latest available calibration files. We extracted the source spectra and light curves from a source-centered circular regions with a radius of 20\arcsec\ for all three instruments. The background likewise was extracted from source-free regions of the same radius in the same chips. We note that there are no MOS1 data available for observation ObsID 0302970601.

\subsection{\ch\ observations}
\label{chandradata}

\source\ was observed by the \ch\ advanced CCD Imaging Spectrometer (ACIS) several times in 2002 and 2009 (see Table~\ref{tab:observations}) providing a total exposure time of  55.4 ks. In all observations the source is located in ACIS-S3 except for the observation ObsID 10512 in which the detector ACIS-I3 was used. Following the standard pipeline procedure,\footnote{\url{https://cxc.cfa.harvard.edu/ciao/threads/index.html}} we reprocessed the data to extract new event files (level 2) using the task {\sc chandra$\_$repro} from the software package {\sc ciao} v4.12 with an up-to-date {\sc caldb} v4.9.1. We then extract the source and background spectra from circular regions with a radii of 10$\arcsec$ and 30$\arcsec$, respectively.

\begin{figure}
\begin{center} 
\includegraphics[width=0.9\columnwidth]{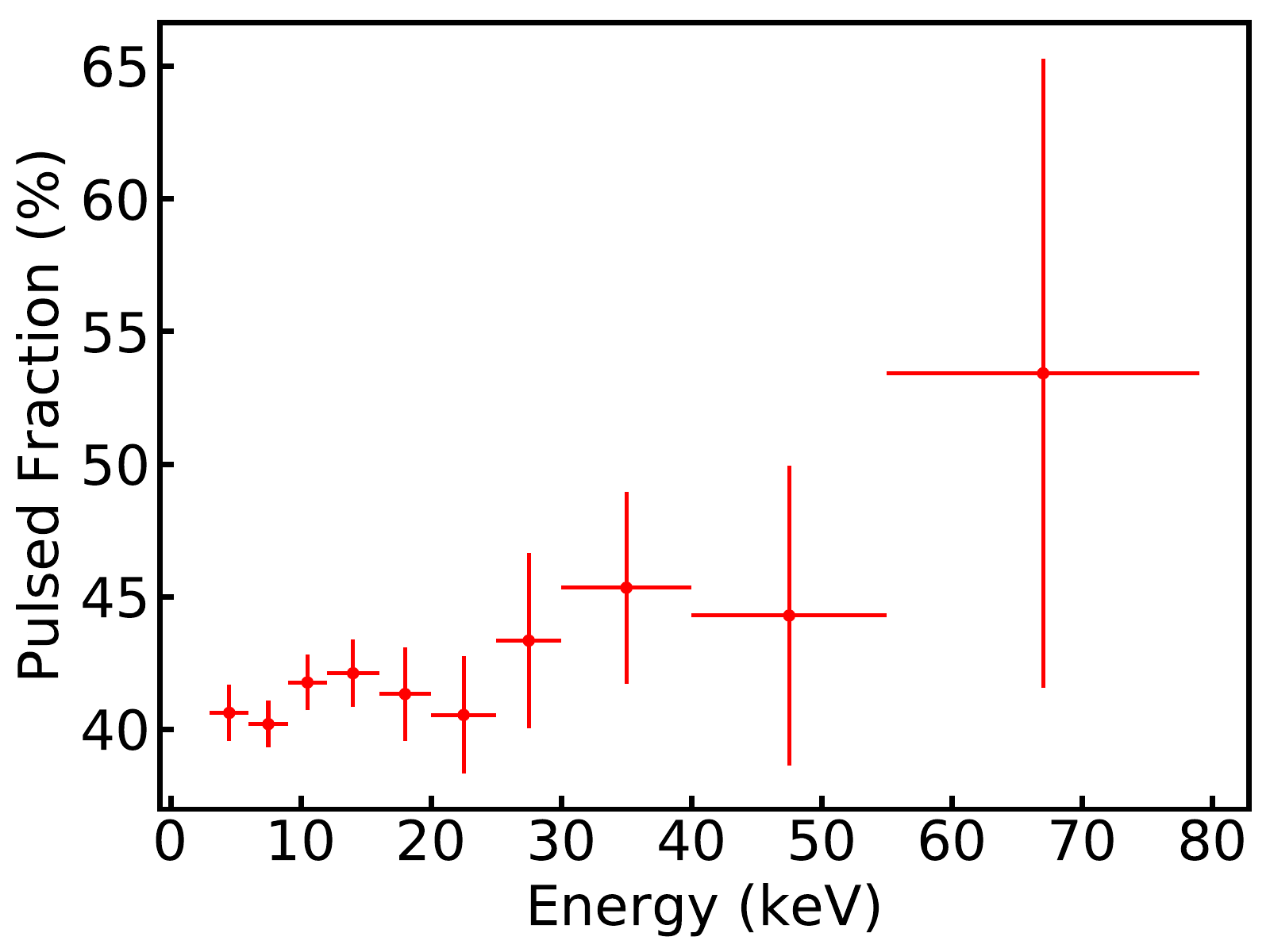}
\end{center}
\caption{Energy dependence of the pulse fraction of \source\ obtained from the \Nu\ observation.}
\label{fig:PF} 
\end{figure}

\subsection{UKIDSS/UKIRT observations}
\label{ukidssdata}
In order to study the type of the companion star in \source\ using the methods explained in \cite{Karasev2015} and \cite{Nabizadeh2019}, the magnitudes of the star in two near-IR filters $H$ and $K$ should be known. We took the magnitude of the counterpart in $K$ filter from the latest public release of the UKIDSS catalog {\it UKIDSS/GPS DR11 PLUS}\footnote{\url{http://surveys.roe.ac.uk/wsa/}}. However, the magnitude of the source in the $H$ filter is not present in that catalog. To solve this problem, we performed an additional photometric analysis of UKIDSS image data (id 4543927 observed on 2006 June 12) using PSF-photometry (DAOPHOT II\footnote{\url{http://www.star.bris.ac.uk/~mbt/daophot/}}) methods. 

Having obtained the instrumental magnitudes of all stars in the 3\arcmin{} vicinity of \source\, we were able to compare these instrumental magnitudes with the ones in the standard UKIDSS catalog (HAperMag3). 
We then selected only the stars brighter than 17 magnitudes in the $H$-filter for this analysis, excluding overexposed objects.  Thus we estimated a mean correction value and converted DAOPHOT magnitude (in the $H$-filter) of the probable counterpart into the real/observed magnitude in the corresponding filter (see Table~\ref{tab:2S_IR}). 
We emphasize that \source\ is not detected in the $J$ filter.

\section{Analysis and results}

\begin{table}
	\centering
	\caption{Orbital parameters of \source\citep{Finger1999}.}
	\label{tab:orbital}
	\begin{tabular}{lr}
			\hline \hline

	Orbital period  & 242.18$\pm$0.01 days\\
	$T_{\rm Periastron}$  &  2449616.98$\pm$0.18 JD \\
	$a_{\rm x}\sin i$ & 689$\pm$38 lt-s   \\
	Longitude of periastron & 252.2$\pm$9.4 deg \\
	Eccentricity & 0.8792$\pm$0.0054 \\

		\hline
	\end{tabular}

\end{table}
\subsection{Pulse profile and pulsed fraction}

For timing analysis we used \Nu\ barycentric-corrected and background-subtracted light curves. The binary motion correction was also applied to the light curves to convert the observed time to the binary-corrected time using the orbital parameters, obtained from \citet{Finger1999} given in Table~\ref{tab:orbital}. The long exposure time and high count rate allowed us to determine the spin period of the NS of $P_{\rm spin}$ = 94.7171(3)~s. To obtain the spin period and its uncertainty the standard {\sc efsearch} procedure from the {\sc ftool} package was applied on $10^3$ simulated light curves created by using the count rates and uncertainties of the original 3--79 keV light curve  \citep[see e.g., ][]{Boldin2013}. Considering the wide energy range of \Nu, we were able to study the pulse profile of the source as a function of energy. For this, we first extracted the source and background light curves in five energy bands 3--7, 7--18, 18--30, 30--50 and 50--79 keV. We then combined the light curves extracted from the modules FPMA and FPMB in order to increase the statistics. 

The energy-dependent light curves were folded with the obtained pulse period using the task {\sc efold} from the {\sc xronos} package. Evolution of the pulse profile with energy is shown in the top five panels of Fig.~\ref{fig:PP}. Pulse profiles demonstrate a complicated structure consisting of multiple peaks. The main maximum and the main minimum are around 0.1--0.2 and 0.6--0.7, respectively, where the zero phase is chosen arbitrarily. As can be seen, the pulse profile depends on energy with multi-peak structure becoming more prominent at higher energies. The most significant changes take place around the main minimum and the main maximum of the profile. It can be best illustrated with the hardness ratio constructed using the pulse profiles in 3--7 and 18--30 keV bands as shown in the bottom panel of Fig.~\ref{fig:PP}. The hardness ratio shows two clear hardening of the emission at the rising part of the main maximum and at the center of the main minimum. 

We also calculated the pulsed fraction, determined as $PF = (F_{\max} - F_{\min})/(F_{\max} + F_{\min})$, where the $F_{\max}$ and $F_{\min}$ are the maximum and minimum fluxes of the pulse profile, as a function of energy. In the majority of XRPs, the pulse fraction shows a positive correlation with the energy \citep{Lutovinov2009}, however, as shown in Fig.~\ref{fig:PF}, the pulsed fraction in \source\ has values around 40--50\% with no prominent dependence on the energy.

\begin{table}
	\caption{Phenomenological models used to fit the source spectral continuum.}
	
	\label{tab:model}
	\begin{tabular}{ll} 
		\hline	\hline
		Model & Photon energy distribution \\
		\hline
		\textsc{cutoffpl}       & 	$N(E) = K{E^{-{\rm \alpha}}} {\rm exp}(-E/\beta)$	             \\
		\textsc{po$\times$highecut}          &  $M(E)=K{E^{-{\rm \alpha}}}\exp[(E_{\rm c}-E)/E_{\rm f}]$,    ($E \geq E_{\rm c}$)    \\
		                        &   $M(E) = K{E^{-{\rm \alpha}}}$,       ($E \leq E_{\rm c}$)                    \\

		\textsc{npex}       &  $N(E) = (A_{\rm 1}E^{-\alpha_{\rm 1}}+A_{\rm 2}E^{+\alpha_{\rm 2}})~{\rm exp}(-E/kT)$  	              \\

        \textsc{fdcut}       &  $N(E) = A_{\rm PL} {E^{-\Gamma}}/[{{\rm exp}((E - E_{\rm cut})/E_{\rm fold}) + 1}]$	              \\

        \textsc{comptt}       &  Comptonization model from \citet{Titarchuk1994}  	             \\

		 \hline
		 
	\end{tabular}
      \small

\end{table}

\subsection{Phase-averaged spectroscopy}
\label{Phase-averaged}

\begin{figure}
\begin{center} 
\includegraphics[width=0.9\columnwidth]{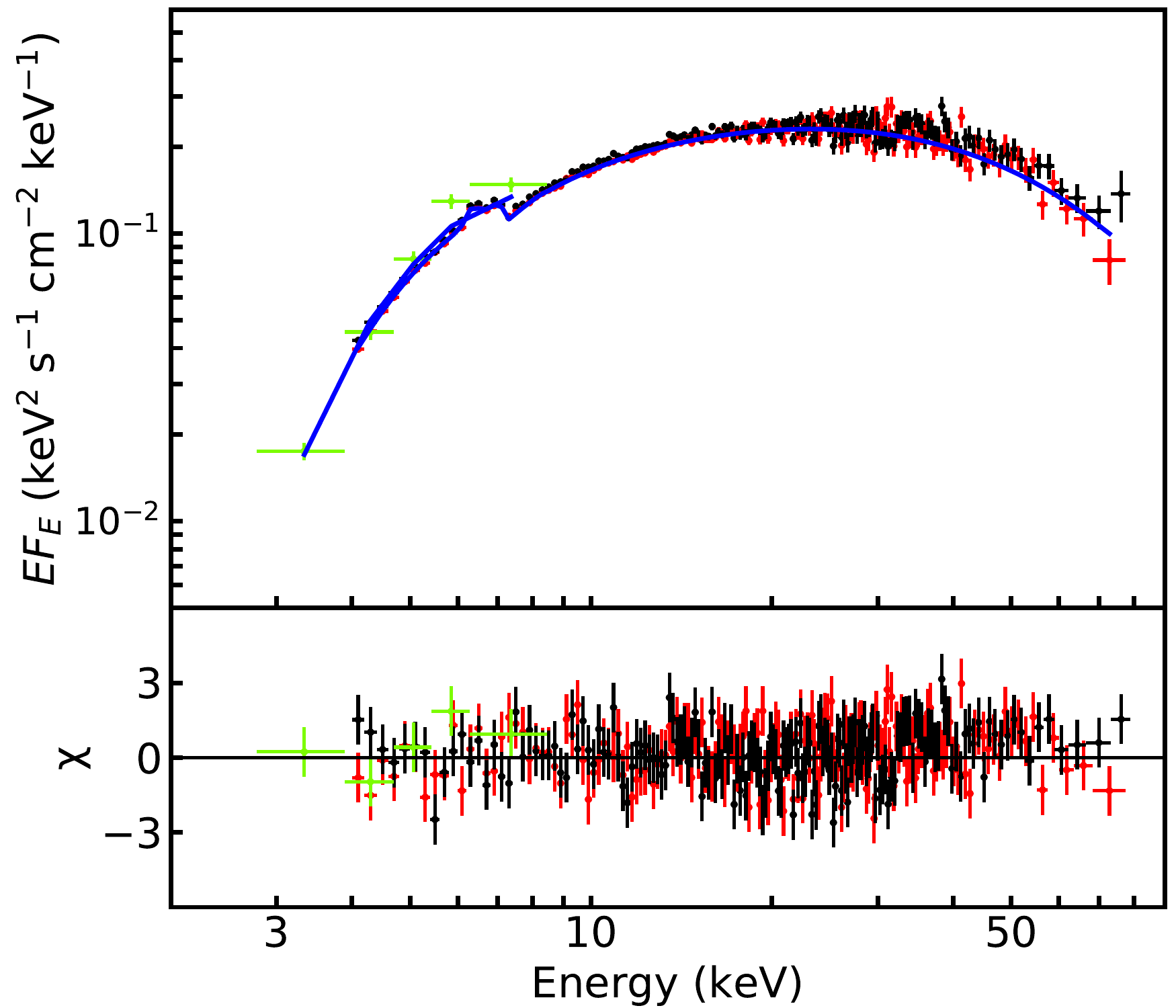}
\end{center}
\caption{\textit{Top panel:} Broad band X-ray spectrum of \source\ extracted from \sw/XRT (green crosses) and  \Nu/FPMA and FPMB (red and black crosses). Solid blue line represents the best-fit model {\sc constant $\times$ tbabs $\times$ (po $\times$ highecut + gau)}. 
\textit{Bottom panel:} Residuals from the best-fit model in units of standard deviations. 
We emphasize that the \sw/XRT spectrum is obtained in the range 0.3--10 keV, however, there are not enough soft X-ray photons below 3~keV because the spectrum is highly absorbed.
}
\label{fig:best-fit} 
\end{figure}

The simultaneous observations of \source\ obtained with \sw/XRT and \Nu\ allowed us to perform the spectral analysis in a broad band, 0.3--79 keV, for the first time for the source. The broadband spectrum of \source\ shown in Fig.~\ref{fig:best-fit} turned out to have a shape typical for XRPs \citep[][]{Filippova2005}. According to \citet{Koyama1990}, the source continuum can be fitted by a phenomenological model such as a power-law with high-energy exponential cut-off. However, to find the best-fit model, we tested several continuum models as listed in Table~\ref{tab:model}. Consequently, the {\sc fdcut} model could not fit the spectrum, while {\sc cutoffpl}, {\sc npex} and {\sc comptt} gave acceptable fits with $\chi^2$ (d.o.f) of 2098 (1769), 1787 (1766) and 2007 (1768), respectively. The model {\sc po $\times$ highecut} fitted the spectrum slightly better with $\chi^2$ (d.o.f) = 1769 (1767). Therefore, and also to be able to make a comparison between our results and the previous studies, we used this preferred model for both the phase-averaged and the phase-resolved analysis. 
The Galactic and intrinsic absorption was modeled using photoelectric absorption model {\sc tbabs} with abundances from \citep{Wilms2000} and atomic cross-sections adopted from \citet{Verner1996}. We also used a Gaussian emission component to account for the narrow fluorescent iron line at 6.4 keV.

The best-fit  composite model ({\sc constant $\times$ tbabs (po $\times$ highecut + gaussian})) along the data and the corresponding residuals are shown in Fig.~\ref{fig:best-fit} and the best-fit parameters and the corresponding uncertainties at 68.3\% (1$\sigma$) confidence level are given in Table~\ref{tab:best-fit}. The fit revealed a large hydrogen column density $N_{\rm H}$ = (22.7$\pm0.7$) $\times 10^{22}$ cm$^{-2}$. We note that the Galactic mean value in the direction to the source is 1.81 $\times 10^{22}$ cm$^{-2}$ \citep{Willingale2013} which is significantly lower than what we have obtained. This discrepancy can be due to a significant intrinsic absorption in the system. To study this, we studied  variations of the column density as a function of orbital phase. 

We utilized the eleven archival observations (see~Table~\ref{tab:observations}) performed at different orbital phases as listed in Table~\ref{tab:nh}. Since the data cover only soft X-ray band below 10 keV, we modeled the spectra using a simple composite model {\sc tbabs $\times$ (po + gaussian)}. We note that the \Nu\ spectra were also fitted using the same model in the energy range 4--10 keV. Due to the lack of high count statistics in some observations we were unable to detect the iron emission line and thus fixed the line centroid energy and width to our best-fit values from the joint \sw+\Nu\ data. Consequently, the column density for different orbital phases are obtained and given in Table~\ref{tab:nh}. The corresponding X-ray flux for each observation was also calculated in the energy range 0.3--10 keV and reported in the same table. The data show strong dependence of $N_{\rm H}$ on the orbital phase as well a correlation with the flux (see~Table~\ref{tab:nh}).  For those observations with lower exposure time, we binned the spectra to have at least 1 count s$^{-1}$ and used W-statistics \citep{Wachter1979} in order to get more reliable fits.

We emphasize that the best-fit model showed no evidence of a Cyclotron Resonant Scattering Feature (CRSF) in the broad-band source spectra (see~Fig. \ref{fig:best-fit}). However, we continued searching for the possible cyclotron line following the steps explained by \citet{Doroshenko2020}. As a result, we did not detect any absorption feature at any energy with significance above $\sim$2.4$\sigma$.

\begin{table}
	\caption{Best-fit parameters for the joint \sw/XRT and \Nu\ phase-averaged spectrum approximated with the {\sc constant $\times$ tbabs(powerlaw $\times$ highecut + gaussian)} model.}
	
	\label{tab:best-fit}
	\begin{tabular}{lccr} 
		\hline	\hline
		Model & Parameters & Unit & Value\\
		\hline
		\textsc{constant}       & 	\Nu$^{a}$	       &                     & 1.015$\pm0.003$ \\
		                        & 	\sw/XRT$^{b}$	       &                     & 0.69$\pm0.03$ \\
		\textsc{tbabs}          &  $N_{\rm H}$ & $10^{22}$ cm$^{-2}$ &  22.7$\pm0.7$ \\

		\textsc{powerlaw}       &  $\Gamma$    &		             &  1.23$\pm0.02$ \\
				                &  norm        & ($\times10^{-2}$)	 & 3.6$\pm0.2$ \\
        \textsc{highecut}       &  $E_{\rm cut}$  & keV	             &  8.2$\pm0.2$ \\
                                &  $E_{\rm fold}$ & keV	             &  28.6$\pm0.8$ \\

        \textsc{gaussian}       &  $E_{\rm Fe}$ &	keV	             &  6.35$\pm0.03$ \\
				&  $\sigma_{\rm Fe}$ & keV	                         &  0.10$^{+0.07}_{-0.09}$ \\
                &  norm & $10^{-4}$ ph s$^{-1}$cm$^{-2}$             &  1.3$\pm0.3$ \\
        \hline
         $F_{0.3-79}$$^{c}$ &	& $10^{-9}$ erg s$^{-1}$cm$^{-2}$           & 1.07$\pm0.01$\\
		 $F_{0.3-10}$$^{c}$ & & $10^{-10}$ erg s$^{-1}$cm$^{-2}$           & 4.10$\pm0.09$ \\
		 \hline

		 $\chi^2$   & 	&	& 1769 \\
		 d.o.f.   & 	&	& 1767  \\
		 \hline
		 
	\end{tabular}
      \small
\tablefoot{    
\tablefoottext{a}{Cross-calibration normalization constant between \Nu/FPMA and FMPB. }
\tablefoottext{b}{Cross-calibration normalization constant between  \Nu/FPMA and \sw/XRT.}
\tablefoottext{c}{Unabsorbed X-ray flux.}
}
\end{table}

\begin{figure}
\begin{center}
\includegraphics[width=0.9\columnwidth]{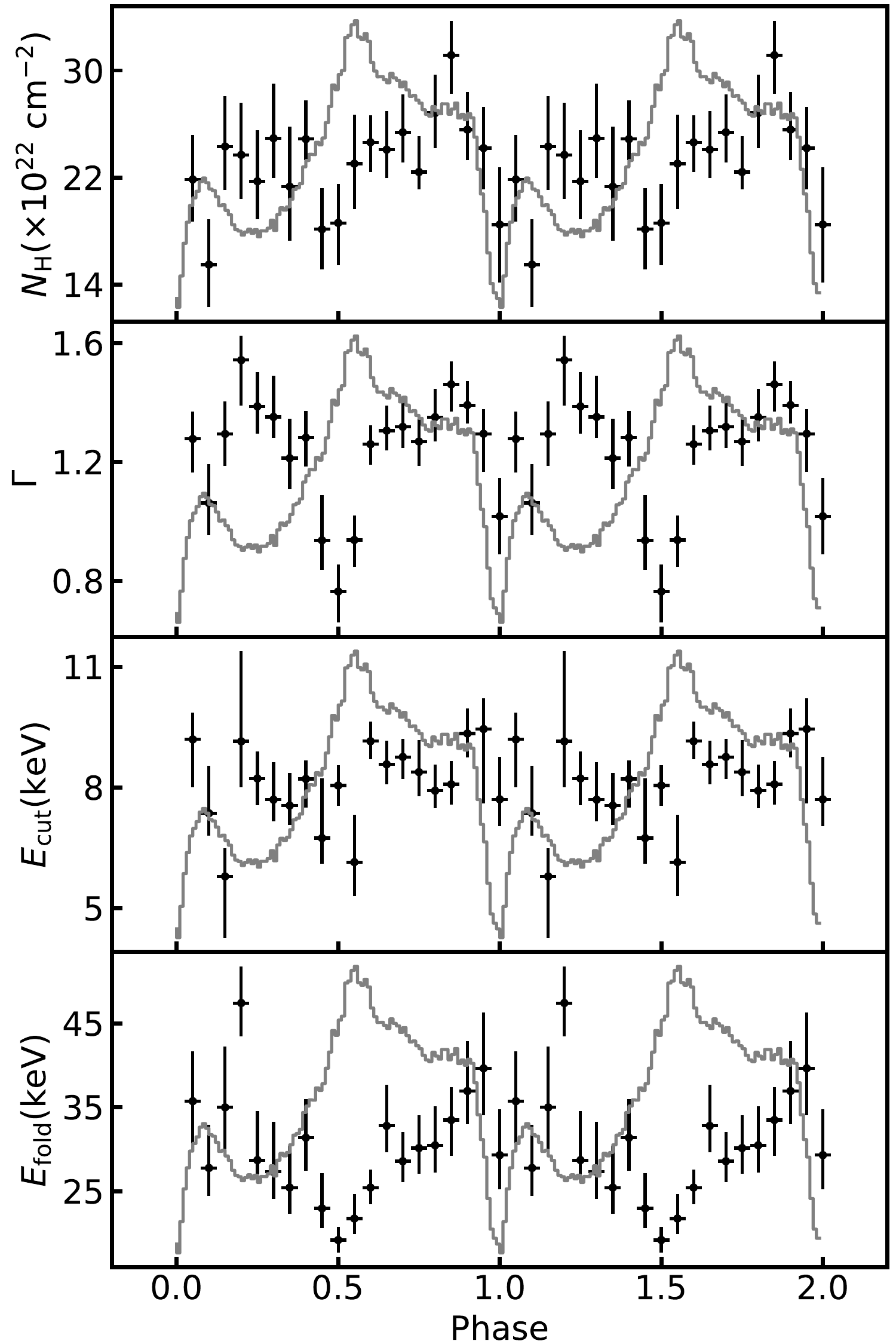}
\end{center}
\caption{Variations of the spectral parameters of the best-fit model as a function of pulse phase. The black crosses from the uppermost to the lowest panel show: neutral hydrogen column density $N_{\rm H}$ in units of $10^{22}$ cm$^{-2}$, photon index, cutoff energy, folding energy. The full energy (3--79 keV) averaged pulse profile of the source is shown in gray in each panel. Errors are 1$\sigma$.
}

\label{fig:phase-res} 
\end{figure}

\begin{table*}
    \centering
    \caption{Spectral parameters of \source\ as a function of orbital phase.}
    \label{tab:nh}

    \begin{tabular}{lccccc}
    \hline  \hline
    Observatory & ObsID & Orbital phase & $\Gamma$ & $N_{\rm H}$  &  Flux$^{a}$  \\
     & & & & ($\times 10^{22}$ cm$^{-2}$)  & (\ergscm)\\ 
    \hline
    \ch\ & 2691        & 0.003 & 0.08$\pm0.39$ & 52$\pm$11      & 2.4$^{+0.6}_{-0.4} \times 10^{-11}$\\
    \sw\ & 00033739001 & 0.009 & 0.7$^{+1.0}_{-0.6}$    & 68$^{+22}_{-10}$    & 2.5$^{+3.5}_{-0.9} \times 10^{-10}$ \\
    \Nu+\sw$^{b}$ & 90201056002+00088089001     & 0.029 & 1.15$\pm0.03$    & 21.0$\pm0.9$    & 3.14$^{+0.06}_{-0.08} \times 10^{-10}$ \\
    \xm\ & 0302970801  & 0.160 & 0.7$\pm0.3$    & 20$^{+4}_{-3}$      & 1.8$^{+0.3}_{-0.2} \times 10^{-12}$ \\
    \xm\ & 0302970601  & 0.426 & 1.6$^{+0.6}_{-0.5}$    & 22$^{+6}_{-5}$      & 1.7$^{+1.9}_{-0.6} \times 10^{-12}$ \\
    \ch\ & 10512       & 0.748 & 0.4$\pm0.2$    & 9$^{+10}_{-8}$      & 9.6$^{+4.2}_{-3.8} \times 10^{-13}$ \\
    \ch\ & 2692        & 0.924 & $-0.3^{+1.3}_{-0.9}$   & 13$^{+11}_{-7}$     & 9.7$^{+2.4}_{-2.1} \times 10^{-13}$ \\
    \sw\ & 00609139000 & 0.991 & 2.0$\pm0.7$    & 106$^{+18}_{-17}$   & 1.9$^{+5.8}_{-1.0} \times 10^{-9}$ \\
    \sw\ & 00707545000 & 0.992 & $-0.1^{+0.5}_{-0.7}$   & 31$^{+10}_{-8}$      & 4.7$^{+1.0}_{-0.7} \times 10^{-10}$ \\
    \sw\ & 00745966000 & 0.996 & 0.4$\pm0.8$  & 32$^{+14}_{-11}$     & 6.0$^{+3.9}_{-1.4} \times 10^{-10}$ \\

    \hline
    \end{tabular}

    \small
    	\tablefoot{
\tablefoottext{a}{Unabsorbed X-ray fluxes in energy range 0.3--10 keV.}
\tablefoottext{b}{The fit parameters and flux obtained from a joint fit in range 0.3--10 keV.}
}

\end{table*}


\subsection{Phase-resolved spectroscopy}
\label{phase-resolved}

Phase-resolved spectroscopy is a useful technique to study the spatial properties of the emitting region of the NS. Based on the good counting statistics, we extracted twenty equally spaced phase bins (see upper panel in Fig.~\ref{fig:PP}) from the available \Nu\ observation of \source. Each spectrum was fitted with our best-fit model ({\sc constant $\times$ tbabs (po $\times$ highecut + gaussian}); see Sec.~\ref{Phase-averaged}). Similar to the phase-average spectral analysis, we fixed the iron line width at 0.1 keV for all 20 spectra. The evolution of the fit parameters are shown in Fig.~\ref{fig:phase-res}.

The hydrogen column density $N_{\rm H}$ varies in the range of (15--31) $\times 10^{22}$ cm$^{-2}$ showing a marginally significant deviation from a constant. The photon index $\Gamma$ shows a similar behavior as $N_{\rm H}$ varying from $\sim$0.7 at the main maximum to $\sim$1.5 at the second minimum of the pulse. The cutoff energy $E_{\rm cut}$ remains almost constant around 8 keV throughout the pulse with variations between 5.8 and 9.5 keV. The folding energy $E_{\rm fold}$ is more variable reaching $\sim$48 keV near the second minimum of the pulse and decreasing down to 19 keV at the main maximum. 

\begin{figure}
\begin{center} 
\includegraphics[width=0.9\columnwidth]{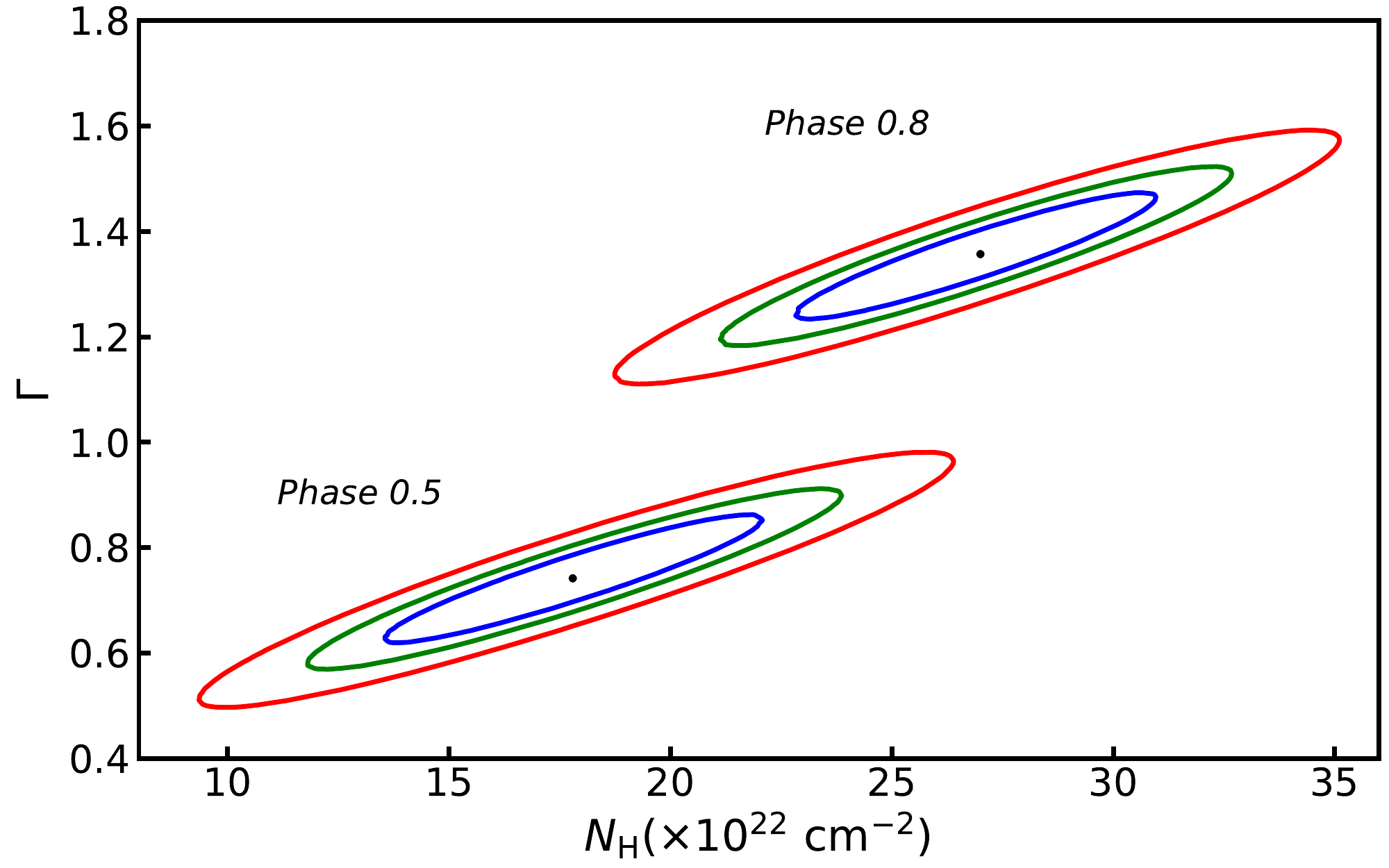}
\end{center}
\caption{Confidence contours of $N_{\rm H}$ versus $\Gamma$ obtained using the best-fit model for the spin phase-resolved spectra at phases 0.5 and 0.8 (see the text). The blue, green and red contours correspond to the 1$\sigma$, 2$\sigma$ and 3$\sigma$ confidence levels obtained using $\chi^2$ statistics for 2 free parameters.
}
\label{fig:contours} 
\end{figure}

Because there is possible strong internal correlation between $N_{\rm H}$ and $\Gamma$ in soft X-ray band, we constructed the confidence contour plot of these two parameters using the spectra of the phases 0.5 and 0.8 where these parameters have different values (see Fig.~\ref{fig:contours}). We see that although the values of $N_{\rm H}$ for two phases agree within 2$\sigma$ confidence level, the photon index is significantly different pointing to the intrinsic variability of the spectrum.

\subsection{X-ray position and IR companion}

Due to the poor localization of \source, the nature of the optical counterpart in this system is yet unknown. \source\ is located in the Scutum region which is crowded by transient XRPs and their companions \citep{Koyama1990Galactic-arm}. In order to determine the source position from the X-ray data, we selected one of the \ch\ observation (ObsID 2689). Using the task {\sc celldetect} standard routines,\footnote{\url{https://cxc.cfa.harvard.edu/ciao/threads/celldetect/}} we obtained the source position at R.A. = 18$^{\rm h}48^{\rm m}16\fs8$ and Dec. = $-2\degr25\arcmin25\farcs1$ (J2000).
A total uncertainty of 1$\arcsec$ (at 90\% confidence level radius), including the systematic uncertainty of \ch\ absolute positions,\footnote{\url{https://cxc.harvard.edu/cal/ASPECT/celmon/}} was obtained for the localization accuracy of the source. 

We also obtained the astrometrically corrected source coordinates from the averaged image of all available \sw/XRT observations using the online XRT products generator.\footnote{\url{https://www.swift.ac.uk/user_objects/}} Based on this, the source is located at R.A. = 18$^{\rm h}48^{\rm m}16\fs91$ and Dec. = $-2\degr25\arcmin26\farcs1$ (J2000) with an error radius of $2\farcs5$ at 90\% confidence level, which is fully consistent with the \ch\ results (see Fig.~\ref{fig:counterpart}).

\subsection{Nature of IR companion}

\begin{figure}
\begin{center}
\includegraphics[width=\columnwidth,trim={0.5cm 10cm 0.5cm 10cm},clip]{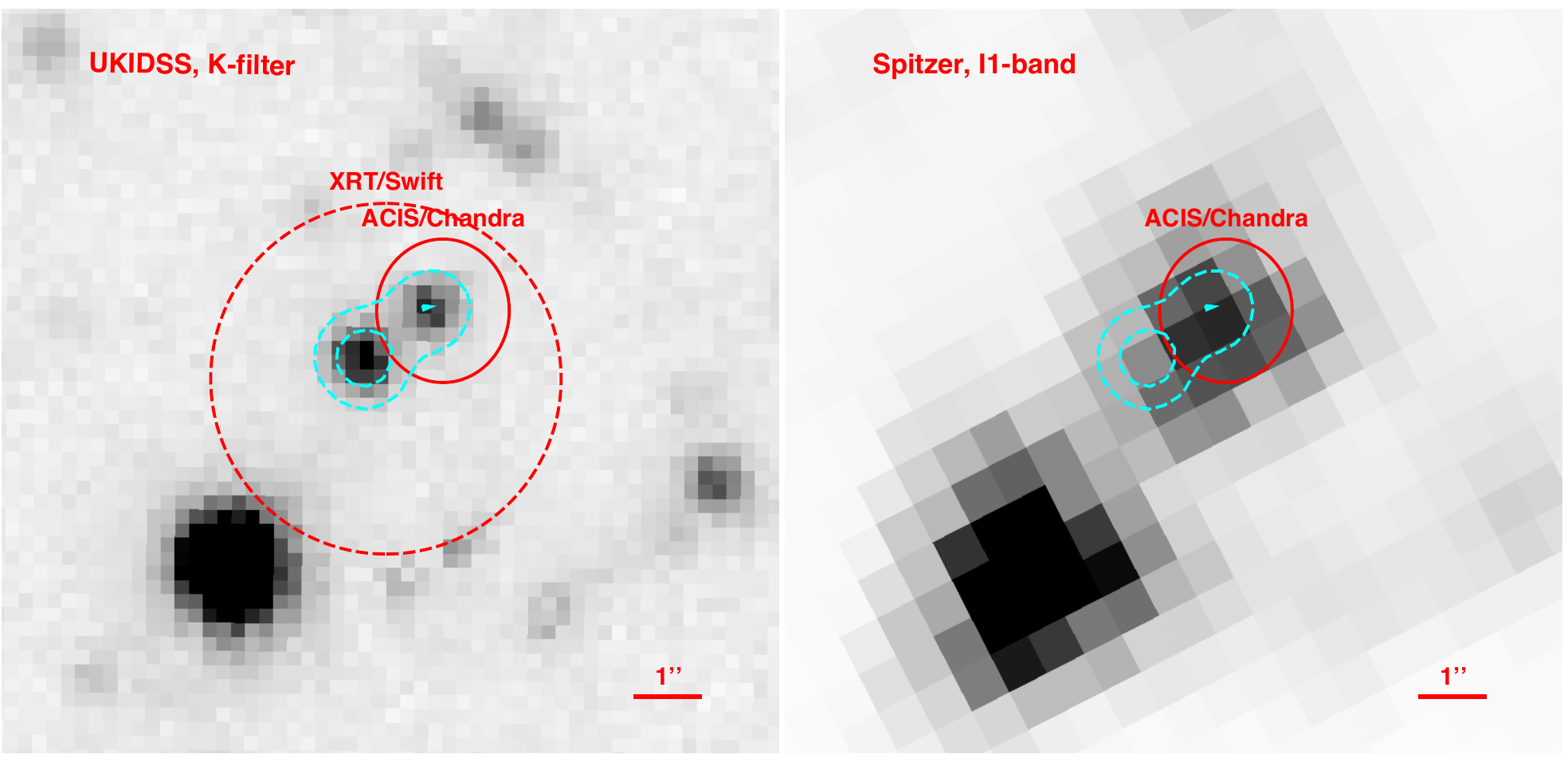}
\end{center}
\caption{Images of the sky around \source\ in the $K$-filter obtained by the UKIRT-telescope (GPS/UKIDSS sky survey, left) and in the $3.6\mu$-band obtained by the {\it Spitzer} telescope (right). The red circles indicate an uncertainty for the source position based on the {\it Swift} (dashed line) and {\it Chandra} (solid line) data, respectively. Cyan contours mark two IR objects closest to the X-ray position.
}
\label{fig:counterpart} 
\end{figure}

Using the results of \ch\ localization and data of the UKIDSS near-IR sky survey, we were able to identify the IR-counterpart of \source\ (see Fig.~\ref{fig:counterpart}, left panel). The coordinates and magnitudes of the IR counterpart are given in Table~\ref{tab:2S_IR}. An expected class of the star as well as the distance to it can be estimated using the method successfully applied earlier in a number of sources \citep[see, e.g.,][]{Karasev2015,Nabizadeh2019}.

\begin{table}
	\centering
	\caption{Coordinates and IR-magnitudes of the counterpart of \source\ based on {UKIDSS/GPS} and \textit{Spitzer} data.}
	\label{tab:2S_IR}
	\begin{tabular}{lr}
			\hline \hline
	RA & 18$^{\rm h}$48$^{\rm m}$16\fs87 \\ 
	Dec & -02$\degr25\arcmin25\farcs2$ \\ 
	$l$ & 30\fdg4151 \\
	$b$ &  $-$0\fdg4031 \\
	$H$ &  $17.82\pm0.04$  \\
	$K$ & $15.52\pm0.03$ \\
	$[3.6]$ $\si{\micro\metre}$ & $12.74\pm0.07$ \\
	$[4.5]$ $\si{\micro\metre}$ & $12.35\pm0.14$  \\
	$[5.4]$ $\si{\micro\metre}$ & $11.66\pm0.11$ \\ 

		\hline
	\end{tabular}
\end{table}

\begin{figure}
\centering 
\includegraphics[width=0.9\columnwidth]{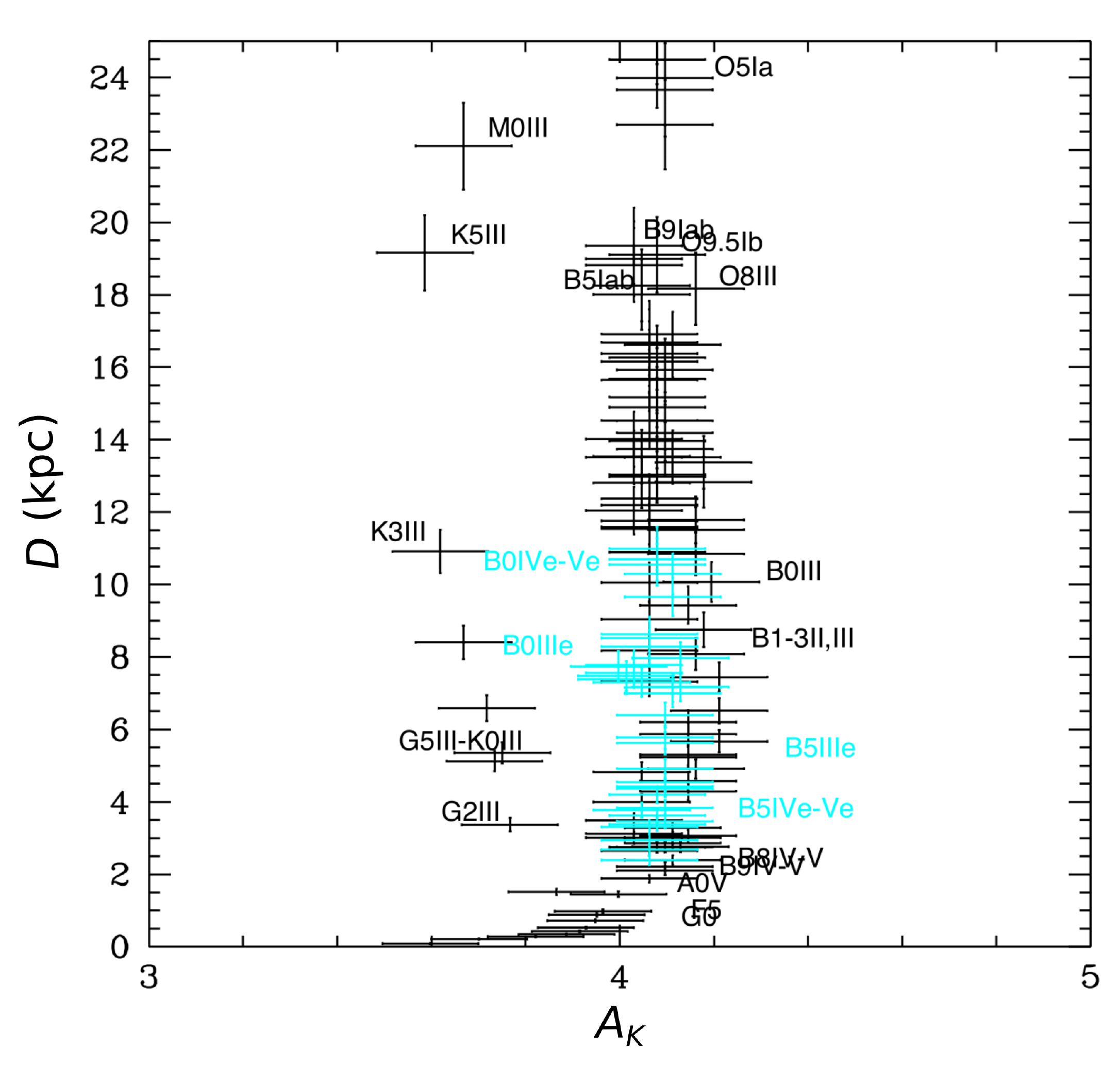}
\caption{`Distance-extinction' diagram showing how far the star (black dots for normal and cyan for Be ones) of a specific class should be located if it is a counterpart of \source\ and the appropriate extinction towards such a star.}
\label{fig:IR_class} 
\end{figure}

Comparing the measured color of the source  $(H-K)=2.30\pm0.05$ with intrinsic colors $(H-K)_0$ of different classes of stars \citep[][all values were converted into the UKIRT filter system via relations from \citealt{Carpenter2001}]{2014AcA....64..261W, 2015AN....336..159W}, we can estimate corresponding extinction corrections $E(H-K)=(H-K)-(H-K)_0$. \source\ is located far from the Galactic bulge, therefore, we can use a standard extinction law \citep{Cardelli1989} to transform each $E(H-K)$ into the extinction $A_{K}$. In turn, comparing absolute magnitudes of the same classes of stars $M_{\rm K}$  \citep{2000MNRAS.319..771W, 2006MNRAS.371..185W,2007MNRAS.374.1549W} with the measured magnitude of the source in the $K$-filter, we are able to estimate a probable distance $D$ to each class of stars as $5-5\log_{10}D=M_{\rm K} - K + A_K$. Results of this approach are indicated in Fig.~\ref{fig:IR_class}.

Unfortunately, having magnitudes only in two filters makes it challenging to come up with a solid conclusion about the nature of the IR companion, however, the extinction $A_ K$ towards the system can be roughly estimated. According to Fig.~\ref{fig:IR_class}, $A_ K\simeq4.1$ accounts for OB-stars, including giants or supergiants, and $A_ K\simeq3.7$ for red giants. By converting these extinction magnitudes into the hydrogen column density $N_{\rm H}$ using the standard relations $A_V=8.93\times A_K$ \citep{Cardelli1989} and $N_{\rm H} = 2.87\times 10^{21} \times A_V$ \citep{Foight2016}, we obtain $N_{\rm H} \simeq(10-11)\times10^{22}$ cm$^{-2}$ for different types of the companion stars. At the same time, the X-ray spectrum revealed a significantly higher column density of 22.7 $\times$  10$^{22}$ cm$^{-2}$, that is typical for highly absorbed HMXB systems \citep[see, e.g.,][]{Rahoui2008}. This circumstance may indicate that \source\ belongs to this class of binary systems.

To clarify the nature of the companion, we also used the mid-IR data obtained by {\it Spitzer} telescope\footnote{\url{http://www.astro.wisc.edu/sirtf/}} (see Table~\ref{tab:2S_IR}). However, as can be seen from Fig.~\ref{fig:counterpart} there is another star located near the probable IR-counterpart of \source. Spatial resolution of {\it Spitzer} did not allow us to fully resolve these objects (see cyan contours in Fig.~\ref{fig:counterpart}). Therefore, we were not able to exclude that the resulting mid-IR fluxes mentioned in Table~\ref{tab:2S_IR} are affected by the confusion of these two stars.

Nevertheless, if we assume OB-supergiant (B9Iab, B5Iab, O5Ia etc.) as a counterpart of \source, the distance to the source is expected to be more than $\sim$16 kpc (see Fig.~\ref{fig:IR_class}). This is in line with \citet{Koyama1990} who estimated a 10-kpc distance to the source based on the high $N_{\rm H}$ value in the source spectrum. Our spectral analysis also supports these results as $N_{\rm H}$ shows variations on the orbital timescale from $\sim$(1--2) $\times$ 10$^{23}$ cm$^{-2}$ at phase around 0.5 to $\sim$10$^{24}$ cm$^{-2}$ around the periastron passage. The lowest value of $N_{\rm H}$ is almost an order of magnitude higher than the Galactic mean value in the direction to the source. This fact along with a positive correlation of the $N_{\rm H}$ value with the X-ray flux points to the presence of a strong stellar wind in the system. Similar behavior is observed in other XRPs with hypergiant optical companions \citep[e.g., for GX 301--2,][]{2014MNRAS.441.2539I}. But at the same time, we cannot rule out other classes of stars to be a companion. Thus, to establish reliably the nature of the IR companion of \source\ the spectroscopic observations in the near-IR band, for example K-band spectroscopy, are required. After the class of the companion star will be established we will be able to use the diagram shown in Fig.~\ref{fig:IR_class} to estimate the distance to the source with high accuracy.

\section{Discussion and Conclusions}
\label{discussion}

In this work, we presented the results of the detailed X-ray and IR analysis of the poorly studied XRP \source\ and its companion during the type I outburst of the source in 2017. For X-ray analysis, we used a single \Nu\ observation performed during the outburst and several X-ray observations obtained by \xm, \ch\ and \sw. For IR analysis, data obtained from UKIDSS/GPS and $\it Spitzer$/GLIMPSE surveys were used. 

In order to determine the magnetic field strength of the NS in the system which was one of our prime goals, we searched for possible cyclotron absorption line in the broad-band \Nu\ spectrum. Such feature was not discovered in phase-averaged nor in phase-resolved spectra of \source. Therefore, it can be inferred that either the line does not exist in the considered energy range or it is too weak to be detected with the current sensitivity of the observations. In the former case, considering the lower and upper limits of the operating energy-band of the \Nu\ instruments, we only can estimate the magnetic field strength of the source to be either weaker than ~$\sim$4 $\times$ 10$^{11}$ G or stronger than $\sim$7 $\times$ 10$^{12}$ G. Further sensitive observations are required to make a solid conclusion.

In order to determine the nature of the companion and the distance to \source, we performed analysis of the IR data. However, the availability of the magnitudes only in two ($H$ and $K$) filters  allowed us to roughly classified the IR-companion in \source\ as an OB-supergiant star located at a distances of more than  $\sim$16 kpc. To establish a more accurate estimation for the nature of the IR-companion in this system as well as the distance to the source, sensitive spectroscopic observations in the near-IR band (i.e. K-band spectroscopy) are required. 
Our conclusion about the class of the optical companion is supported by the X-ray spectral properties of the source. 
The good coverage of the binary orbit with observations in the soft X-rays, allowed us to investigate the variation of column density $N_{\rm H}$ as a function of orbital phase which revealed the presence of a strong stellar wind in the system. However, we emphasize that an extensive study of the iron line are required to support this interpretation \citep[see][]{2014MNRAS.441.2539I}.

The estimation of the distance to \source\ can be also done using the observed fluxes and presumable luminosity of the source in the different states. Particularly, in the low state, when the observed flux drops down to about $10^{-12}$ \flux\ (see Table~\ref{tab:nh}), one can expect the luminosity of the source to be above $\sim$10$^{34}$ erg s$^{-1}$ in the case of the ongoing accretion \citep{Tsygankov2017cold-disk,Tsygankov2019cold-disk} and, therefore, the distance to the system can not be below $\sim$10~kpc. From another side, the peak luminosity during type I outbursts from the transient XRPs can be of the order of 10$^{37}$ erg s$^{-1}$. Taking into account the maximal observed flux from \source\ of around $10^{-9}$ \flux\ one can estimate an upper limit on the distance as $\sim$15~kpc. These rough estimates agree with results obtained from the IR data.

\begin{acknowledgements}
This work was supported by the grant 14.W03.31.0021 of the Ministry of Science and Higher Education of the Russian Federation. We also acknowledge the support from the Finnish Cultural Foundation through project number 00200764 and 85201677 (AN), the Academy of Finland travel grants 317552, 322779, 324550, 331951, and 333112, the National Natural Science Foundation of China grants 1217030159, 11733009, U2038101, U1938103, and the Guangdong Major Project of the Basic and Applied Basic Research grant 2019B030302001 (LJ). This work is based in part on data of the UKIRT Infrared Deep Sky Survey. Also, part of this work is based on observations made with the Spitzer Space Telescope, which is operated by the Jet Propulsion Laboratory, California Institute of Technology under a contract with NASA. 
\end{acknowledgements}

\bibliographystyle{aa}
\bibliography{library}

\end{document}